\documentclass[journal, onecolumn]{IEEEtran}

\usepackage{cite}
\usepackage{version}
\usepackage{fullpage,epsfig,graphicx,psfrag,color,subfig}
\usepackage{amsmath}
\usepackage{amssymb}

\usepackage{xspace}
\usepackage{mathrsfs}
\usepackage{amsopn}

\def\Var{\mathop{\rm Var}\nolimits}%
\newcommand{\Ac}{\mathcal{A}}

\newcommand{\Cc}{\mathcal{C}}

\newcommand{\Ec}{\mathcal{E}}

\newcommand{\Sc}{\mathcal{S}}

\newcommand{\Xc}{\mathcal{X}}
\newcommand{\Yc}{\mathcal{Y}}

\newcommand{\pen}{{P_e^{(n)}}}

\newcommand{\aep}{{\mathcal{T}_{\epsilon}^{(n)}}}
\newcommand{\aepvar}{{\mathcal{T}_{\epsilon'}^{(n)}}}

\newcommand{\Mh}{{\hat{M}}}

\newcommand{\lh}{{\hat{l}}}
\newcommand{\mh}{{\hat{m}}}

\newcommand{\Ut}{{\tilde{U}}}
\newcommand{\Vt}{{\tilde{V}}}

\newcommand{\lt}{{\tilde{l}}}

\newcommand{\ut}{{\tilde{u}}}
\newcommand{\vt}{{\tilde{v}}}

\def\e{\epsilon}

\DeclareMathOperator\E{E}
\let\P\relax
\DeclareMathOperator\P{P}

\DeclareMathOperator\C{C}

\newcommand{\Bern}{\mathrm{Bern}}

\def\textiid{i.i.d.\@\xspace}
\newcommand\iid{\ifmmode\text{ i.i.d. } \else \textiid \fi}

\newtheorem{theorem}{Theorem}

\newtheorem{remark}{Remark}

\newcommand{\qedsymbol}{$\square$}

\begin{document}
\title{An Achievability Scheme for the Compound Channel with State Noncausally Available at the Encoder}
\author{Chandra Nair\IEEEauthorrefmark{1}, Abbas El Gamal\IEEEauthorrefmark{2} and Yeow-Khiang Chia\IEEEauthorrefmark{2}
\thanks{\IEEEauthorrefmark{1} Chandra Nair is with the Chinese University of Hong Kong} \thanks{\IEEEauthorrefmark{2} Abbas El Gamal and Yeow-Khiang Chia are with Stanford University}%
\thanks{This work was partially supported by the Institute of Network Coding \big(formed using a grant from the University Grants Committee of the Hong Kong Special Administrative Region, China (Project No. AoE/E-02/08)\big), and by the Institute of Theoretical Computer Science and Communication, both at the Chinese University of Hong Kong.}
}

\maketitle

\begin{abstract}
A new achievability scheme for the compound channel with discrete memoryless (DM) state noncausally available at the encoder is established. Achievability is proved using superposition coding, Marton coding, joint typicality encoding, and indirect decoding. The scheme is shown to achieve strictly higher rate than the straightforward extension of the Gelfand-Pinsker coding scheme for a single DMC with DM state, and is optimal for some classes of channels.
\end{abstract}

\section{Introduction} \label{sect:1}
\begin{figure}[h]
\begin{center}
\psfrag{A}[c]{$M$}
\psfrag{B}[c]{$$}
\psfrag{C}[c]{$X^n$}
\psfrag{D}[l]{$$}
\psfrag{E}[c]{}
\psfrag{F}[c]{$Y^n_{1}$}
\psfrag{F2}[c]{$Y^n_{2}$}
\psfrag{G}[c]{$\Mh_1$}
\psfrag{G2}[c]{$\Mh_2$}
\psfrag{P1}[c]{\rm Encoder}
\psfrag{P4}[c]{\rm Decoder 1}
\psfrag{P5}[c]{\rm Decoder 2}
\psfrag{P2}[c]{$p(s)$}
\psfrag{P3}[c]{$p(y_1,y_2|x,s)$}
\psfrag{s}[b]{$S^n$}
\includegraphics[width=0.7\linewidth]{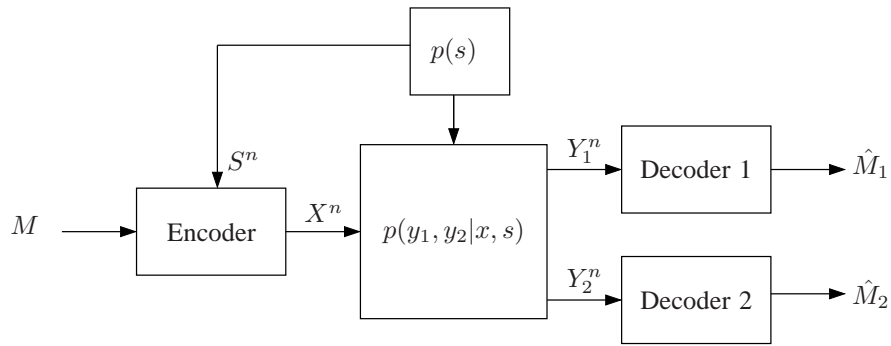} 
\caption{Sending common message over DM-BC with DM state.} \label{fig1}
\end{center}
\end{figure}

Consider the problem of reliable communication over a compound channel with discrete memoryless (DM) state, where a sender wishes to communicate a message to a receiver with the state sequence available noncausually at the encoder. For simplicity we consider the case when the compound channel comprises only two discrete memoryless channels (DMCs) with DM state. This setup is essentially the same as sending a common message over a 2-receiver discrete memoryless broadcast channel (DM-BC) with DM state when the state in available noncausally at the encoder as shown in Figure~\ref{fig1}. As such, we focus our discussion throughout the paper on this equivalent setup.

The capacity for the single receiver case, widely referred to as the Gelfand--Pinsker channel, was established in~\cite{Gelfand} as
\[
C_{\rm GP} = \max_{p(u|s),\, x(u,s)} (I(U:Y) - I(U;S)).
\]
The proof of achievability involves randomly generating a subcodebook for each message. To send a message, the sender finds a codeword in the message subcodebook that is jointly typical with the given state sequence. The receiver decodes the codeword and hence finds the message. The details of the proof can be found, for example, in~\cite[Lecture 7]{El-Gamal--Kim2010}. 

A straightforward extension of this Gelfand--Pinsker scheme to the DM-BC with DM state yields the lower bound on capacity 
\begin{equation}\label{GF}
C \ge \max_{p(u|s), \, x(u,s)}\min\{I(U;Y_1) - I(U;S), I(U;Y_2) - I(U;S)\}.
\end{equation}
In~\cite{pablo}, it is conjectured that this rate is optimal in general. We show that this is not the case. 
We devise a new coding scheme for this channel that involves superposition coding, Marton coding, joint typicality encoding, and indirect decoding~\cite{Nair}. Our scheme yields the following lower bound on capacity.

\begin{theorem} \label{thm:1}
The common message capacity of the DM-BC with state information available non-causally at the sender is lower bounded by
\begin{align*}
C \ge \max \min & \{ I(W, U; Y_1) - I(W,U;S), \; I(W, V; Y_2) - I(W,V;S), \\
&  \quad \frac{1}{2}\left(I(W, U; Y_1) - I(W,U;S)+ I(W, V; Y_2) - I(W,V;S) - I(U;V|W,S)\right) \},
\end{align*}
where the maximization is over distributions $p(w,u,v|s)$ and functions $x(w,u,v,s)$.
\end{theorem}

It is easy to see that this lower bound is at least as large as~\ref{GF}. We simply set $U=V=\emptyset$. We will show that our lower bound can in fact be strictly larger than~\ref{GF}.

In the following section, we formally define the problem of sending a common message over a DM-BC with DM state and describe the new coding scheme. In section~\ref{sect:3}, we show through an example that the new lower bound can be strictly larger than the straightforward extension of the Gelfand-Pinsker result. In section~\ref{sect:4}, we present several classes of channels for which the new rate is optimum, including a class of compound Gaussian channels where the new rate achieves the dirty paper coding rate~\cite{costa} for both channels simultaneously.

The notation used in this paper will follow that of El Gamal--Kim Lecture Notes on Network Information Theory~\cite[Lecture 1]{El-Gamal--Kim2010}.
\section{Achievability Scheme} \label{sect:2} 
Consider a 2-receiver DM-BC with DM state $(\Xc, \Sc, \{p(y_1,y_2|x,s)p(s), \Yc_1,\Yc_2)$ consisting of a finite input alphabet $\Xc$, finite output alphabets $\Yc_1,\Yc_2$, a finite state alphabet $\Sc$, two a collection of conditional pmfs $p(y_1,y_2|x,s)$ on $\Yc_1\times \Yc_2$, and a pmf $p(s)$ on the state alphabet $\Sc$.  

A $(2^{nR}, n)$ code for the DM-BC with noncausal state information available at the encoder consists of: (i) a message set $[1:2^{nR}]$, (ii) an encoder that assigns a codeword $x^n(m, s^n)$ to each message $m$ and state sequence $s^n$, and (iii) two decoders, decoder 1 assigns an estimate $\hat{m}_1(y_1^n) \in [1:2^{nR}]$ or an error message $\mathrm{e}$ to each received sequence $y^n_1$ and decoder 2 that assigns an estimate $\hat{m}_2(y_2^n) \in [1:2^{nR}]$ or an error message $\mathrm{e}$ to each received sequence $y_2^n$. We assume that $M$ is uniformly distributed over $[1:2^{nR}]$. The probability of error is defined as $\pen = \P\{\Mh_1 \neq M \text{ or }  \Mh_2 \neq M\}$.

A rate $R$ is said to be achievable if there exists a sequence of $(2^{nR}, n)$ codes with $\pen \to 0$ as $n\to \infty$. The capacity $C$ is the supremum of all achievable rates.
\medskip

The main result in this paper is the lower bound on the common message capacity of the DM-BC with DM state available non-causally at the encoder in Theorem~\ref{thm:1}. The proof of this theorem follows.

\subsection*{Codebook generation}
\begin{itemize}
\item For each $m$, generate $2^{n{T_0}}$ $w^n(m,l_0)$ sequences according to $\prod_{i=1}^n p_{W}(w_i)$.
\item For each $(m,l_0)$ pair, generate $2^{nT_1}$ $u^n(m,l_0,l_1)$ sequences according to $\prod_{i=1}^n p_{U|W}(u_i|w_i)$.
\item For each $(m,l_0)$ pair, generate $2^{nT_2}$ $v^n(m,l_0,l_2)$ sequences according to $\prod_{i=1}^n p_{V|W}(v_i|w_i)$.
\end{itemize}

\subsection*{Encoding}
\begin{figure}[!h]
\begin{center}
\psfrag{u}[c]{$U^n$}
\psfrag{v}[c]{$V^n$}
\psfrag{u1}[l]{$u^n(1,1)$}
\psfrag{u2}[c]{$u^n(1,2)$}
\psfrag{ur}[l]{$u^n(1,2^{nT_1})$}
\psfrag{vr}[c]{$v^n(1,2^{nT_2})$}
\psfrag{v1}[c]{$v^n(1,1)$}
\psfrag{v2}[c]{$v^n(1,2)$}
\psfrag{aep}[c]{$(w^n, s^n, u^n, v^n) \in \aep$}
\psfrag{c1}[c]{$\Cc(1)$}
\psfrag{c2}[c]{$\Cc(2)$}
\psfrag{cr}[c]{$\Cc(2^{nR})$}
\scalebox{0.8}{\includegraphics{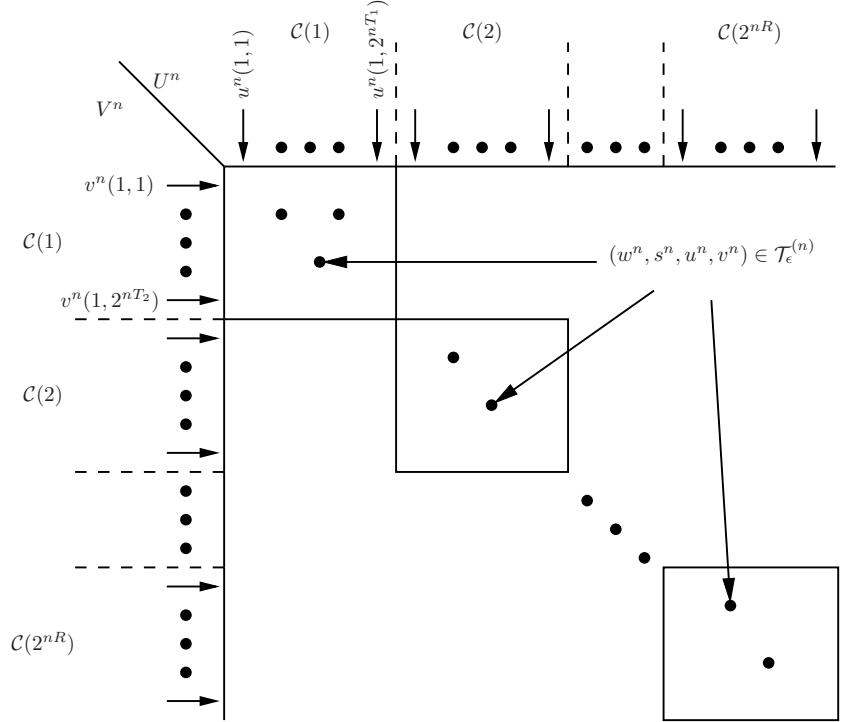}}
\end{center}
\caption{Achievability scheme.} \label{fig:2}
\end{figure}

The encoding procedure is illustrated in Figure~\ref{fig:2}.
\begin{itemize}
\item Given message $m$ and state sequence $s^n$, the encoder finds $l_0 \in [1:2^{nT_0}]$ such that $(w^n(l_0),s^n) \in \aep$. If there is more than one $l_0$, it chooses the smallest one. If there is none, it chooses $l_0=1$.
\item The encoder next finds $l_1 \in [1:2^{nT_1}]$ and $l_2 \in [1:2^{nT_2}]$ such that \\ $(w^n(m,l_0), s^n, u^n(m,l_0, l_1), v^n(m,l_0, l_2)) \in \aep$. If there is more than one such pair, it chooses the pair with the smallest indices, first in $l_1$, then in $l_2$. If there is none, it chooses $(1, 1)$.
\item The encoder transmits $x(w_i, u_i, v_i, s_i)$ for $i \in [1:n]$.
\end{itemize}
Note that this scheme is essentially Marton coding with only diagonal product bins. Interestingly, the same encoding scheme can be used if we wish to send a common message $M_0$ to both receivers and private messages $M_1$ to $Y_1$ and $M_2$ to $Y_2$. 

\subsection*{Decoding}
Let $\e' > \e>0$.
\begin{itemize}
\item Decoder 1 finds $m$ indirectly by decoding $(m,l_0)$. It declares that $\mh_1$ is sent if it is the unique message such that $(w^n(\mh_1, \lh_0), u^n(\mh_1, \lh_0, \lh_1), y_1^n) \in \aepvar$ for some $\lh_0 \in [1:2^{nT_0}], \; \lh_1 \in [1:2^{nT_1}]$.
\item Decoder 2 finds $m$ indirectly by decoding $(m,l_0)$. It declares that $\mh_2$ is sent if it is the unique message such that $(w^n(\mh_2,\lh_0), v^n(\mh_2, \lh_0, \lh_2), y_2^n) \in \aepvar$ for some $\lh_0 \in [1:2^{nT_0}], \; \lh_2 \in [1:2^{nT_2}]$.
\end{itemize}

\subsection*{Analysis of probability of error} 
An error may occur if either the encoder does not find a quadruple such that $(w^n(m,l_0), s^n, u^n(m,l_0, l_1),$\\$ v^n(m,l_0, l_2)) \in \aep$, or there is an error made by decoder 1 or 2.  

We now analyze the probability of error averaged over codebooks. Without loss of generality, assume  $M=1$ is sent and $(L_0,L_1,L_2)$ are the corresponding indices. Define the encoding error events
\begin{align*}
\Ec_{01} &= \{(S^n, W^n(1,l_0))\notin \aep  \text{ for all } l_0\},\\
\Ec_{02} &= \{(S^n, W^n(1,L_0), U^n(1, L_0, l_1), V^n(1, L_0,l_2))\notin \aep \text{ for all } l_1, l_2\}
\end{align*}
Then the total encoding error probability is
\[
\P(\Ec_0) \le \P(\Ec_{01}) + \P(\Ec_{02} \cap \Ec_{01}^c).
\]
By the covering lemma~\cite[Lecture 3]{El-Gamal--Kim2010}, 
the first term $\P(\Ec_{01})\to 0$ as $n \to \infty$ if
\begin{align*}
T_0 > I(W;S).
\end{align*}
Next, consider the second probability of error term
{\allowdisplaybreaks
\begin{align*}
\P(\Ec_{02} \cap \Ec_{01}^c) &= \P\{(S^n, W^n(1,L_0), U^n(1,L_0, l_1), V^n(1,L_0, l_2)) \notin \aep\ \mbox{ for all } l_1, l_2\} \\
& \le \sum_{(w^n, s^n) \in \aep(W,S)}\P\{W^n(1,L_0) = w^n,S^n = s^n\}\P\{\Ec_{02}(s^n,w^n)\},
\end{align*}
where $\Ec_{02}(s^n,w^n)$ denotes the event that $\{(S^n=s^n, W^n(1,l_0)=w^n, U^n(1, L_0, l_1), V^n(1, L_0,l_2))\notin \aep\}$ for all $l_1$ and $l_2$, conditioned on the fact that the pair $(w^n, s^n) \in \aep(W,S)$.
}

We show in Appendix~\ref{appen:1} that $\P(\Ec_{02}(s^n,w^n)) \to 0$ as $n \to \infty$ if
\begin{align*}
T_1 &> I(U;S|W)+\delta(\e), \\
T_2 &> I(V;S|W)+\delta(\e),  \\
T_1 + T_2 &> I(U;S|W) + I(V;S|W) + I(U;V|W,S) +\delta(\e).
\end{align*}

Next consider the probability of decoding error. Consider the following error events for decoder 1
\begin{align*}
\Ec_{11} &= \{(S^n, W^n(1,L_0), U^n(1,L_0, L_1), Y_1^n) \notin \aepvar\},\\
\Ec_{12} &= \{(S^n, W^n(m, \lt_0), U^n(m, \lt_0, \lt_1), Y_1^n) \in \aepvar \text{ for some } \lt_0 \in [1:2^{nT_0}],\; \lt_1 \in [1:2^{nT_1}],\; m \neq 1\}.
\end{align*}

The probability of error restricted to $\Ec_{01}^c$ for decoder 1 is upper bounded as
\[
\P(\Ec_1) \leq \P(\Ec_{11} \cap \Ec_{01}^c) + \P(\Ec_{12}).
\]

By the law of large numbers, the second term $\P(\Ec_{11}\cap\Ec_{01}^c) \to 0$ as $n\to \infty$. By the packing lemma~\cite[Lecture 3]{El-Gamal--Kim2010}, the third term $\P(\Ec_{12}) \to 0$ as $n\to \infty$ if 
\begin{align*}
R+ T_0 + T_1 < I(W,U;Y_1)- \delta(\e).
\end{align*}

Similarly, the probability of error at decoder 2 tends to zero as $n\to \infty$ if
\begin{align*}
R+ T_0 + T_2 < I(W,V;Y_2) - \delta(\e).
\end{align*}
Thus the overall probability of error tends to zero as $n \to \infty$ if 
\begin{align*}
R+ T_0 + T_1 &< I(W,U;Y_1), \\
R+ T_0 + T_2 &< I(W,V;Y_2), \\
T_0 &> I(W;S), \\
T_1 &> I(U;S|W), \\
T_2 &> I(V;S|W),  \\
T_1 + T_2 &> I(U;S|W) + I(V;S|W) + I(U;V|W,S). 
\end{align*}
Performing Fourier-Motzkin Elimination on the stated rate constraints then gives the achievable rate stated in Theorem~\ref{thm:1}. \qedsymbol
\medskip

\noindent{\em Remarks}:
 
\begin{enumerate}
\item It suffices to set $X$ as a deterministic function of $W$ and $S$ in \eqref{GF} and in Theorem 1.  In \eqref{GF}, if $X$ is a probabilistic mapping of $(W,S)$, by the functional representation lemma~\cite{El-Gamal--Kim2010} it can always be expressed as a function of $(W,S,Q)$, where $Q$ is independent of $(W,S)$. Defining $W' = (W, Q)$, we obtain $X =x(W',S)$, $I(W';Y_1) - I(W';S) \ge I(W;Y_1) - I(W;S)$ and $I(W';Y_2) - I(W';S) \ge I(W;Y_2) - I(W;S)$. Similar reasoning can also be applied to Theorem 1.

\item Theorem 1 can be readily extended to any finite number of receivers (equivalently, compound channel comprising a finite number of DMCs with DM state). In this case we have the common auxiliary random variable $W$ and as many individual auxiliary random variables as the number of receivers.
\end{enumerate}
\section{Example} \label{sect:3}
\begin{figure}[!h]
\begin{center}
\psfrag{s0}[c]{$S=0$}
\psfrag{s1}[c]{$S=1$}
\psfrag{x}[c]{$X$}
\psfrag{y}[c]{$Y_1$}
\psfrag{z}[c]{$Y_2$}
\psfrag{0}[c]{$0$}
\psfrag{1}[c]{$1$}
\scalebox{0.8}{\includegraphics{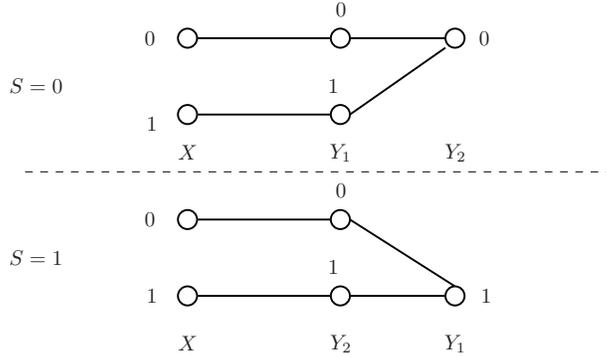}}
\end{center}
\caption{Example DM-BC with DM state.} \label{fig:3}
\end{figure}

We now show through the example in Figure~\ref{fig:3} that the achievable rate in Theorem~\ref{thm:1} can be strictly larger than the rate achievable by the straightforward extension of the Gelfand-Pinsker coding scheme to the 2-receivers DM-BC with state given in~\ref{GF}, which we denote by $R_{\rm GP}$.  

We have $|\Xc| = |\Yc_1| = |\Yc_2| = |S| = 2$ and $\P\{S=0\} = 1/2$. The top half of the example corresponds to the channel transition probabilities when $S=0$ while the bottom half corresponds to the channel transition probabilities when $S=1$. 

From Theorem~\ref{thm:1}, we set $W=\emptyset$, $U = Y_1$, $V = Y_2$ and $\P\{X=0|S=0\} = \P\{X=0|S=1\} = 0.5$. It is easy to verify that this choice of auxiliary random variables gives us an achievable rate of $R = 0.5$. It is also clear that $C \le I(X;Y|S) = H(Y|S) = 0.5$. Therefore, Theorem~\ref{thm:1} achieves the common message capacity for this example.

\subsection{$R_{\rm GP} < C$}
 Expanding $I(U;Y_1) - I(U;S)$ in~\ref{GF}, we obtain
\begin{align*}
I(U;Y_1) - I(U;S) &= I(U;Y_1,S) - I(U;S) - I(U;S|Y_1) \\
& = H(Y_1|S) - H(Y_1|U,S) - I(U;S|Y_1) \\
& \le H(Y_1|S) \le \frac{1}{2}.
\end{align*}
To achieve $R_{\rm GP} = H(Y_1|S)$, we require that $U\to Y_1\to S$ form a Markov chain and $Y_1$ a function of $(U,S)$. Since $Y_1 = X$ when $S=0$, we require that $X$ is a function of $U$ when $S=0$. Similarly, from $I(U;Y_2) - I(U;S)$, we require $U\to Y_2\to S$ and $Y_2$ a function of $(U,S)$. This implies that $X$ is a function of $(U,S)$. To further achieve $R_{\rm GP} = 0.5$, we require that $\P\{X=0|S=0\} = \P\{X=0|S=1\} = 0.5$. 

Let 
\begin{align*}
\P\{U=i|X=0, S=0\} = a_i, \\
\P\{U=i|X=1, S=0\} = b_i, \\ 
\P\{U=i|X=0, S=1\} = c_i, \\
\P\{U=i|X=1, S=1\} = d_i.
\end{align*}

Since $X$ is a function of $(U,S)$, at least one of the two parameters $a_i$ and $b_i$ is equal to zero  and at least one of $c_i$ and $d_i$ is also equal to zero. Further, from the Markov chain conditions $\P\{U=i|Y_2=0, S=0\} = \P\{U=i|Y_2=0,S=1\}$ and $\P\{U=i|Y_1=1, S=0\} = \P\{U=i|Y_1=1,S=1\}$, we obtain
\begin{align*}
\frac{a_i + b_i}{2} &= c_i, \\
\frac{c_i + d_i}{2} & = b_i.
\end{align*} 

If $a_i= 0$, $b_i = 2c_i$ and $d_i = 3c_i$. Since one of $c_i, d_i =0$, this means that $a_i = b_i = c_i = d_i =0$ or $\P\{U=i\} = 0$, which is a contradiction. Similarly, $b_i = 0$ forces $\P\{U=i\} = 0$, which is again a contradiction. This shows that there is no $U$ with the required properties. Hence, $R_{\rm GP} < C$.

In fact, by means of a symmetrization argument given in Appendix~\ref{appen:2}, we can show that $R_{\rm GP}$ can be computed exactly and is approximately equal to $0.41$, implying a gap of $0.09$ from $C$. 

\section{Special Classes of Channels} \label{sect:4}
Theorem~\ref{thm:1} achieves the common message capacity in the following cases.
\subsection{A class of deterministic channels with state}
If both $Y_1$ and $Y_2$ are functions of $(X,S)$ and $I(Y_1;Y_2|S) = 0$, then
\begin{align*}
C = \max_{p(x|s)}\min\{ H(Y_1|S), H(Y_2|S)\}.
\end{align*}
The example given in Section~\ref{sect:3} belongs to this class of channels. Achievability follows from Theorem~\ref{thm:1} by setting $W = \emptyset$, $U = Y_1$ and $V = Y_2$. The converse follows from the  fact that $C \leq  \max_{p(x|s)}\min\{ I(X;Y_1|S), I(X;Y_2|S)\}$. 

\begin{remark}
One can also generalize this result to the class where $Y_1$ and $Y_2$ are functions of $(X,S)$; $Y_1$ and $Y_2$ share common information (in the sense of G\"{a}cs-K\"{o}rner), i.e. there exists $ Z=f(Y_1) = g(Y_2)$, and further $I(Y_1;Y_2|S,Z) = 0$. The achievability follows from Theorem~\ref{thm:1} by setting $W = Z$, $U = Y_1$ and $V = Y_2$. 
\end{remark}

\subsection{A class of compound Gaussian channels}
We now develop a Gaussian analog of the example in Section~\ref{sect:4}. Let $S = (T, Z_S)$ where $T \sim \Bern(\alpha)$ and $Z_S \sim N(0, Q_{T})$. The channel is defined as follows. When $T = 0$, we have
\begin{align*}
Y_1 &= g_1X + Z_S + Z_{1}, \\
Y_2 & = 0,
\end{align*}
where $Z_{1} \sim N(0, 1)$. When $T=1$, we have
\begin{align*}
Y_1 &= 0, \\
Y_2 & = g_2X + Z_S + Z_{2},
\end{align*}
where $Z_{2} \sim N(0, 1)$. The random variables $(T,Z_S),Z_1,Z_2$ are mutually independent. Since $Z_S \sim N(0, Q_{T})$, we may have different variances in different states. Further, we assume an average transmit power constraint: $\sum_{i=1}^n \E(x_{i}^2(m, S^n)) \le nP, \; m \in [1:2^{nR}]$. 

An upper bound on the capacity of this channel is 
\begin{align*}
C \le \max_{p(x|s):\, \E(X^2) \le P} \min\{I(X;Y_1|S), I(X;Y_2|S)\}.
\end{align*}

It is easy to show that $I(X;Y_1|S) \le \alpha \C(g_1^2P_1)$ and $I(X;Y_2|S) \le \bar{\alpha} \C(g_2^2P_2)$, where $\alpha P_1 + \bar{\alpha}P_2 = P$ and $\C(P') = (1/2)\log (1 + P')$. From the writing on dirty paper result~\cite{costa}, in the single state case, the rate is $\C(P)$. Can we achieve the dirty paper coding rate for both $Y_1$ and $Y_2$ \textit{simultaneously} for this more complicated class of compound Gaussian channels?

Using Theorem~\ref{thm:1}, we set $W= T$. When $T=0$, we set
\begin{align*}
U = X_0 + \frac{g_1P_1}{1+g_1P_1}Z_S, \mbox{ and }
V = T,
\end{align*}
where $X_1 \sim N(0, P_1)$. When $T=1$, we set
\begin{align*}
U = T, \mbox{ and }
V = X_1 + \frac{g_2P_2}{1+g_2P_2}Z_S,
\end{align*}
where $X_1 \sim N(0, P_2)$ and $\alpha P_1 + \bar{\alpha}P_2  = P$. This choice of random variables gives us the following achievable rate
\begin{align*}
R &< I(T;Y_1) + I(U;Y_1|T) - I(U;Z_S|T) - H(T), \\
R &< I(T;Y_2) + I(V;Y_2|T) - I(V;Z_S|T) - H(T), \\
2R &< I(T;Y_1) + I(U;Y_1|T) - I(U;Z_S|T) - H(T) + \\
&\quad I(T;Y_2) + I(V;Y_2|T) - I(V;Z_S|T) - H(T) + I(U;V|T,Z_S).
\end{align*}
Since $I(T;Y_1) = I(T;Y_2) = H(T)$ and $I(U;V|T,Z_S) = 0$, simplifying the expression gives us
\begin{align*}
R < \max_{\alpha P_1 + \bar{\alpha}P_2 = P} \min\{\alpha C(g_1^2 P_1), \bar{\alpha}C(g_2^2 P_2)\},
\end{align*}
which shows that we can achieve the dirty paper coding rate for both channels simultaneously.
\section{Conclusion}
We established a new achievable rate for the compound channel with DM state available noncausally at the encoder. The new achievable rate is shown to be strictly larger than the straightforward extension of the Gelfand-Pinsker coding scheme for a single state case. This result also implies that the straightforward extension of the Gelfand-Pinsker coding scheme for transmission over a DM-BC with DM state is not optimum.

\section*{Acknowledgements}
The authors wish to thank Tsachy Weissman for bringing this problem to their attention.

\bibliographystyle{IEEEtran}
\bibliography{gp}
\appendices
\section{Bounding $\P(\Ec_{02}(s^n,w^n))$} \label{appen:1}
The technique we use for bounding the term $\P(\Ec_{02}(s^n,w^n))$ is similar to that in the proof of the mutual covering lemma in~\cite[Lecture 9]{El-Gamal--Kim2010}.

$\P(\Ec_{02}(s^n, w^n))$ is given by the probability of the event: $\{s^n, w^n, U^n(\lt_1), V^n(\lt_2)) \notin \aep\}$ for all $\lt_1 \in [1:2^{nT_1}]$ and $\lt_2 \in [1:2^{nT_2}]$; where $U^n(\lt_1)$ and $V^n(\lt_2)$ are independently generated, conditioned on the given $w^n$, according to $\prod_{i=1}^n{p_{U|W}(u_i|w_i)}$ and $\prod_{i=1}^n{p_{V|W}(v_i|w_i)}$ respectively. Note that we are given  $(s^n, w^n) \in \aep$.

To show that $\P(\Ec_{02}) \to 0$ as $n \to \infty$, let $\Ac = \{(\lt_1,\lt_2) : (s^n, w^n, \Ut^n(\lt_1), \Vt^n(\lt_2)) \in \aep\}$ and $I(\lt_1,\lt_2) = 1$ if $(s^n, w^n, \Ut^n(\lt_1), \Vt^n(\lt_2)) \in \aep$ and $0$ otherwise. Then, $|\Ac| = \sum_{\lt_1,\lt_2} I(\lt_1,\lt_2)$ and the expected number of jointly typical sequences is given by {\allowdisplaybreaks
\begin{align*}
\E |\Ac| = \sum_{\lt_1,\lt_2} \P\{(s^n, w^n, \Ut^n(\lt_1), \Vt^n(\lt_2)) \in \aep\}.
\end{align*}
We further have the following bound on the probability:
\begin{align*}
& \P\{(s^n, w^n, \Ut^n(\lt_1), \Vt^n(\lt_2)) \in \aep\} \\
&= \sum_{\ut^n \in \aep(\Ut|w^n,s^n)} p(\ut^n)\P\{(s^n, w^n, \Ut^n(\lt_1), \Vt^n(\lt_2)) \in \aep|\Ut^n(\lt_1) =\ut^n\} \\
&= \sum_{\ut^n \in \aep(\Ut|w^n,s^n)} \prod_{i=1}^n p_{U|W}(\ut_i|w_i)\P\{(s^n, w^n, \ut^n, \Vt^n(\lt_2)) \in \aep\} \\
& \stackrel{.}{=} \sum_{u^n \in \aep(U|w^n,s^n)} 2^{-nH(U|W)}2^{-nI(S,U;V|W)} \\
& \stackrel{.}{=} 2^{-n(I(U;S|W) + I(S;V|W) + I(U;V|W,S))}.
\end{align*} }
Hence, we have
\begin{align*}
\E|\Ac| \ge 2^{n(T_1+T_2)}2^{-n(I(U;S|W) + I(S;V|W) + I(U;V|W,S) + \delta(\e))}.
\end{align*}
Next, let{\allowdisplaybreaks
\begin{align*}
p_1 &= \P\{(s^n, w^n, \Ut^n(1), \Vt^n(1)) \in \aep\}, \\
p_2 &= \P\{(s^n, w^n, \Ut^n(1), \Vt^n(1)) \in \aep, (s^n, w^n, \Ut^n(1), \Vt^n(2)) \in \aep\} \\
& = \sum_{\ut^n \in \aep(U|w^n,s^n)}p(\ut^n)\P\{(s^n, w^n, \ut^n, \Vt^n(1)) \in \aep\}\P\{(s^n, w^n, \ut^n, \Vt^n(2)) \in \aep\} \\
& \le 2^{-n(I(U;S|W) + 2I(V;S|W) + 2I(U;V|W,S) - \delta(\e))}, \\
p_3 &= \P\{(s^n, w^n, \Ut^n(1), \Vt^n(1)) \in \aep, (s^n, w^n, \Ut^n(2), \Vt^n(1)) \in \aep\} \\
& = \sum_{\vt^n \in \aep(U|w^n,s^n)}p(\vt^n)\P\{(s^n, w^n, \vt^n, \Ut^n(1)) \in \aep\}\P\{(s^n, w^n, \vt^n, \Ut^n(2)) \in \aep\} \\
& \le 2^{-n(I(V;S|W) + 2I(U;S|W) + 2I(U;V|W,S) - \delta(\e))}, \\
p_4 &= \P\{(s^n, w^n, \Ut^n(1), \Vt^n(1)) \in \aep, (s^n, w^n, \Ut^n(2), \Vt^n(2)) \in \aep\} \\
& = p_1^2.
\end{align*}}
Note that $\E|\Ac| = 2^{n(T_1 + T_2)}p_1$. 
\begin{align*}
\E|\Ac|^2 = 2^{n(T_1 + T_2)}p_1 + \sum_{\lt_1,\lt_2}\sum_{\lt_2 \neq \lt_2'}p_2 + \sum_{\lt_1,\lt_2}\sum_{\lt_1\neq \lt_1'}p_3 + \sum_{\lt_1,\lt_2}\sum_{\lt_1 \neq \lt_1'}\sum_{\lt_2 \neq \lt_2'}p_4.
\end{align*}
Hence, 
\begin{align*}
\Var(|\Ac|) \le 2^{n(T_1 + 2T_2)}p_2 + 2^{n(2T_1 + T_2)}p_3 + 2^{n(T_1 + T_2)}p_1.
\end{align*}
By Chebychev's inequality, we have
\begin{align*}
\P\{|\Ac| =0\} &\le \P\{(|\Ac| - \E|\Ac|)^2 \ge (\E|\Ac|)^2\} \\
& \le \frac{\Var(|\Ac|)}{(\E|\Ac|)^2} \\
& \le 2^{-n(T_1 - I(U;S|W) - \delta(\e))} + 2^{-n(T_2 - I(V;S|W) - \delta(\e))} \\
& \quad + 2^{-n(T_1+T_2 - I(U;S|W) - I(V;S|W) - I(U;V|W,S) - \delta(\e))}
\end{align*}
Hence, $\P\{|\Ac| = 0\} \to 0$ as $n\to \infty$ if the following conditions are satisfied
\begin{align*}
T_1 & > I(U;S|W) + \delta(\e) \\
T_2 &> I(V;S|W) + \delta(\e) \\
T_1+T_2 &> I(U;S|W) + I(V;S|W) + I(U;V|W,S) + \delta(\e).
\end{align*}

Hence $\P(\Ec_{02}(s^n,w^n))$ goes to $0$ as $n \to \infty$, provided the above conditions are satisfied. 

\section{Exact evaluation of $R_{\rm GP}$} \label{appen:2}
In this apendix, we evaluate $R_{\rm GP}$ using a symmetrization argument. Consider any $(U,S,X)$ defined by $\P\{U=i,S=0\} = u_i,\P\{U=i,S=1\}=v_i,\P\{X=0|U=i,S=0\}=a_i,\P\{X=0|U=i,S=1\}=1-b_i$. From the fact that it suffices to look at $X=f(U,S)$, we have $a_i,b_i \in \{0,1\}$. 

Then the following holds
\[\begin{array}{rclrcl}
H(Y) &= & H \left(\sum_i u_i a_i\right), &\qquad
H(Y|U) &= & \sum_i (u_i + v_i) H\left(\frac{ u_i a_i}{u_i + v_i}\right),\\
H(S)&= &1, &\qquad
H(S|U)&=& \sum_i (u_i + v_i) H\left(\frac{ u_i}{u_i + v_i}\right),\\
H(Z) &=& H\left(\sum_i v_i b_i\right),& \qquad
H(Z|U) &=& \sum_i (u_i + v_i) H\left(\frac{ v_i b_i}{u_i + v_i}\right).
\end{array}\]

Now define a $(U',S,X')$ ($U'$ of size $2|\mathcal{U}|$) according to: \begin{align*}&\P\{U'=(i,1),S=0\} = u_i/2, \P\{U'=(i,2),S=0\}=v_i/2, \\
&\P\{X'=0|U'=(i,1),S=0\}=a_i, \P\{X'=0|U'=(i,2),S=0\}=b_i, \\
&\P\{U'=(i,1),S=1\} = v_i/2, \P\{U'=(i,2),S=1\}= u_i/2, \\
& \P\{X'=0|U'=(i,1),S=1\}=1-b_i, \P\{X'=0|U'=(i,2),S=1\}=b_i.
\end{align*}

Then observe that the new entropies are
\begin{align*}
 H(Y') &= H\left(\sum_i \frac{u_i a_i}{2} + \frac{ v_i b_i}{2} \right) \geq \frac 12 (H(Y) + H(Z)),\\
H(Y'|U') &= \sum_i \frac 12  (u_i + v_i) \left(H\left(\frac{ u_i a_i}{u_i + v_i}\right) +  H\left(\frac{ v_i b_i}{u_i + v_i}\right)\right) = \frac 12 (H(Y|U) + H(Z|U)),\\
H(S)&=1, \\
H(S|U')&= \sum_i (u_i + v_i) H\left(\frac{ u_i}{u_i + v_i}\right) = H(S|U),\\
H(Z') &= h(\sum_i \frac{u_i a_i}{2} + \frac{ v_i b_i}{2} ) \geq \frac 12 (H(Y) + H(Z)),\\
H(Z'|U') &= \sum_i \frac 12  (u_i + v_i)\left( H\left(\frac{ u_i a_i}{u_i + v_i}\right) +  H\left(\frac{ v_i b_i}{u_i + v_i}\right)\right)  = \frac 12 (H(Y|U) + H(Z|U)).
\end{align*}

Thus, $I(U';Y') - I(U';S) = I(U';Z') - I(U';S) \geq \frac 12 \left( I(U;Y) - I(U;S) + I(U;Z) - I(U;S) \right).$ 

\subsection{Maximization of $I(U';Y') - I(U';S)$}
Our maximization problem reduces to maximizing
\begin{align*}
 I(U';Y') - I(U';S)
\end{align*}
over all pmfs with the stated $U'$ structure.
That is, we wish to  maximize
\begin{align*}
 & H\left(\sum_i \frac{u_i a_i}{2} + \frac{ v_i b_i}{2} \right) - \sum_i \frac 12  (u_i + v_i) \left( H\left(\frac{ u_i a_i}{u_i + v_i}\right) +  H\left(\frac{ v_i b_i}{u_i + v_i}\right)\right) \\
&\quad  - 1 + \sum_i (u_i + v_i) H\left(\frac{ u_i}{u_i + v_i}\right)
\end{align*}
subject to $ \sum_i u_i = 0.5, \sum_i v_i = 0.5, a_i,b_i \in \{0,1\}$. The term 
can be rewritten  as
\begin{align*}
 & H\left(\sum_i \frac{u_i a_i}{2} + \frac{ v_i b_i}{2} \right) + \sum_i \frac 12  (u_i + v_i) \left(H\left(\frac{ u_i}{u_i + v_i}\right) -  H\left(\frac{ u_i a_i}{u_i + v_i}\right)\right) \\ \\
&\quad  - 1 + \sum_i \frac 12  (u_i + v_i) \left( H\left(\frac{ v_i}{u_i + v_i}\right) -   H\left(\frac{ v_i b_i}{u_i + v_i}\right)\right).
\end{align*}
Let $\mathcal{I}$ be the set of indices where $a_i=0$ and $\mathcal{J}$ be the set of indices where $b_i=0$. This implies that on $\mathcal{I}^c$ we have $a_i=1$ and on $\mathcal{J}^c$ we have $b_i=1$.

Thus, we wish to maximize
\begin{align*}
 & H\left(\sum_{i\in \mathcal{I}^c} \frac{u_i }{2} + \sum_{i\in \mathcal{J}^c} \frac{ v_i }{2} \right) + \sum_{i \in \mathcal{I}} \frac 12  (u_i + v_i) H\left(\frac{ u_i}{u_i + v_i}\right)  \\
&\quad  - 1 + \sum_{i\in \mathcal{J}} \frac 12  (u_i + v_i)  H\left(\frac{ v_i}{u_i + v_i}\right).
\end{align*}
subject to $ \sum_i u_i = 0.5, \sum_i v_i = 0.5$.

Define the following:
\begin{align*}
\frac{x_1}{2} & = \sum_{i \in \mathcal{I} \cap\mathcal{J}} u_i, \quad
\frac{y_1}{2}  = \sum_{i \in \mathcal{I} \cap\mathcal{J}} v_i, \\
\frac{x_2}{2}  &= \sum_{i \in \mathcal{I} \cap\mathcal{J}^c} u_i, \quad 
\frac{y_2}{2}  = \sum_{i \in \mathcal{I} \cap\mathcal{J}^c} v_i, \\
\frac{x_3}{2}  &= \sum_{i \in \mathcal{I}^c \cap\mathcal{J}} u_i, \quad 
\frac{y_3}{2}  = \sum_{i \in \mathcal{I}^c \cap\mathcal{J}} v_i, \\
\frac{x_4}{2}  &= \sum_{i \in \mathcal{I}^c \cap\mathcal{J}^c} u_i, \quad 
\frac{y_4}{2}  = \sum_{i \in \mathcal{I}^c \cap\mathcal{J}^c} v_i. 
\end{align*}

Observe that $\sum_i x_i = 1, \sum_i y_i=1$.

We note the following as a consequence of the concavity of the entropy function.
\begin{align*}
 & \sum_{i \in \mathcal{I}} \frac 12  (u_i + v_i) H\left(\frac{ u_i}{u_i + v_i}\right) + \sum_{i\in \mathcal{J}} \frac 12  (u_i + v_i)  H\left(\frac{ v_i}{u_i + v_i}\right) \\
& \quad = \sum_{i \in \mathcal{I} \cap \mathcal{J}}   (u_i + v_i) H\left(\frac{ u_i}{u_i + v_i}\right) + \sum_{i \in \mathcal{I}\cap \mathcal{J}^c} \frac 12  (u_i + v_i) H\left(\frac{ u_i}{u_i + v_i}\right) + \sum_{i \in \mathcal{I}^c \cap \mathcal{J}} \frac 12  (u_i + v_i) H\left(\frac{ u_i}{u_i + v_i}\right) \\
& \quad \leq \frac{x_1 + y_1}{2} H\left(\frac{x_1}{x_1 + y_1}\right) + \frac{x_2 + y_2}{4} H\left(\frac{x_2}{x_2 + y_2}\right) + \frac{x_3 + y_3}{4} H\left(\frac{x_3}{x_3 + y_3}\right).
\end{align*}

Therefore we can upper bound the true maximum by the maximum of 
\begin{align*}
H\left(\frac{x_3+x_4}{4} + \frac{y_2+y_4}{4}\right) + \frac{x_1 + y_1}{2} H\left(\frac{x_1}{x_1 + y_1}\right) + \frac{x_2 + y_2}{4} H\left(\frac{x_2}{x_2 + y_2}\right) + \frac{x_3 + y_3}{4} H\left(\frac{x_3}{x_3 + y_3}\right) - 1,
\end{align*}
subject to $\sum_i x_i = 1, \sum_i y_i=1$ and $x_i, y_i \geq 0$.

\medskip

Now, we relax this maximization to $\sum_i x_i + y_i=2$ and $x_i, y_i \geq 0$.

Define the partial sums $s_1 = x_1 + y_1$, $s_2 = x_2 + y_2$, $s_3 = x_3 + y_3$, and $s_4 = x_4 + y_4$. We re-write the maximization as
\begin{align*}
H\left(\frac{s_4}{4} + \frac{y_2+x_3}{4}\right) + \frac{s_1}{2} H\left(\frac{x_1}{s_1}\right) + \frac{s_2}{4} H\left(\frac{y_2}{s_2}\right) + \frac{s_3}{4} H\left(\frac{x_3}{s_3}\right) - 1,
\end{align*}
subject to $0 \leq x_1 \leq s_1, 0 \leq y_2 \leq s_2, 0 \leq x_3 \leq s_3$ and $\sum_i s_i = 2$. 

Using concavity of entropy, we can bound the maximum of the above expression by the maximum of
\begin{align*}
H\left(\frac{s_4}{4} + \frac{y_2+x_3}{4}\right) + \frac{s_1}{2} H\left(\frac{x_1}{s_1}\right) + \frac{s_2 + s_3}{4} H\left(\frac{y_2 + x_3}{s_2+s_3}\right)  - 1,
\end{align*}
subject to $0 \leq x_1 \leq s_1, 0 \leq y_2 + x_3 \leq s_2 + s_3$ and $\sum_i s_i = 2$.

We first maximize with respect to $x_1$ and $y_2 + x_3$ keeping the $s_i$ terms fixed. Observe that the maximization is separable and it is concave in $x_1$ and $y_2 + x_3$. Hence the maximum occurs when the first derivatives are zero; i.e. $x_1 = \frac{s_1}{2}$ and $\frac{s_4}{4} + \frac{y_2+x_3}{4} = 1 - \frac{y_2 + x_3}{s_2+s_3}.$

The second condition implies that
$$ (y_2 + x_3) (\frac 14 + \frac 1{s_2+s_3}) = 1 - \frac{s_4}{4}, ~\mbox{or}~ y_2 + x_3 = \frac{(4 - s_4)(s_2 + s_3)}{4+s_2 + s_3}. $$

Substituting for the optimal choices of  $x_1, y_2 + x_3$, the maximization reduces to that of
$$ \frac{s_1}{2} + (1 + \frac{s_2 + s_3}{4}) H(\frac{4 - s_4}{4+s_2 + s_3}) - 1,$$
subject to $\sum_i s_i = 2$, and $s_i \geq 0$.

Denote $s_2 + s_3 = t$ and rewrite the maximization as
\begin{align*}
\left(1 + \frac{t}{4}\right)H\left(\frac{4-s_4}{4+t}\right) - \frac{t}{2} - \frac{s_4}{2}
\end{align*}
subject to $0 \leq t, 0 \leq s_4, s_4 + t \leq 2$.

\medskip

We divide into four cases: 
\begin{enumerate}
\item The maximum is achieved at some strictly internal point, i.e. no inequality is tight.
\item The maximum is achieved when $t=0$.
\item The maximum is achieved when $s_4=0$.
\item The maximum is achieved when $t + s_4 = 2$ but neither $t$ or $s_4$ is zero.
\end{enumerate}

It is not difficult to verify that the maximum over all four cases is attained by Case 3, with the setting $t = \frac 43, s_4=0$, and $s_1 = \frac 23$. The maximum value is approximately $0.41$.

\subsection{The maximizing $p(u,s), x(u,s)$}

We now show all the relaxations can be made tight, i.e. there exists a suitable choice of $U$ that achieves the derived bound.

Consider the following $|\mathcal{U}|$ with cardinality $3$ defined according to:
\begin{align*}
& \P\{U=1,S=0\} = \frac 16, \P\{X=0|U=1,S=0\} = 0, \\
&  \P\{U=1,S=1\} = \frac 16, \P\{X=1|U=1,S=1\} = 0, \\
&\P\{U=2,S=0\} = \frac 1{12}, \P\{X=0|U=2,S=0\} = 0, \\
&  \P\{U=2,S=1\} = \frac 14, \P\{X=1|U=2,S=1\} = 1, \\
& \P\{U=3,S=0\} = \frac 14, \P\{X=0|U=3,S=0\} = 1,\\
 &  \P\{U=3,S=1\} = \frac 1{12}, \P\{X=1|U=3,S=1\} = 0.
\end{align*}

For this channel observe that 
\begin{align*}
I(U;Y) - I(U;S) &= I(U;Z) - I(U;S) \\
& = \frac 43 H\left(\frac 34 \right) - \frac 23 \\
& \approx 0.41.
\end{align*}

\end{document}